\documentclass[pra,aps,amsmath,amssymb,superscriptaddress]{revtex4}
\usepackage{graphicx}
\usepackage{color}
\newtheorem{thm}{Theorem}

\newcommand{\be}{\begin{equation}}
\newcommand{\ee}{\end{equation}}

\newcommand{\bex}{\begin{eqnarray}}
\newcommand{\eex}{\end{eqnarray}}
\newcommand{\bmin}{\begin{center}\begin{minipage}{460pt}}
\newcommand{\emin}{\end{minipage}\end{center}}

\begin{document}

\title{Violation of no-signaling in higher-order quantum measure theories}
\author{Karthik S. Joshi}
\affiliation{Raman Research Institute, Bangalore, India}
\author{R. Srikanth} 
\email{srik@rri.res.in}
\affiliation{Poornaprajna Institute of Scientific Research, Bangalore, India}
\affiliation{Raman Research Institute, Bangalore, India}
\author{Urbasi Sinha}
\email{usinha@rri.res.in}
\affiliation{Raman Research Institute, Bangalore, India}
\affiliation{Institute for Quantum Computing, Waterloo, Ontario, Canada}

\begin{abstract}
More  general probability sum-rules  for describing  interference than
found  in  quantum mechanics  (QM)  were  formulated  by Sorkin  in  a
hierarchy of  such rules.  The additivity of  classical measure theory
corresponds  to  the second  sum-rule.   QM  violates  this rule,  but
satisfies the third and higher sum-rules.  This evokes the question of
whether there are physical principles that forbid their violation.  We
show that in a theory that is indistinguishable from quantum mechanics
in  first and  second  order interferences,  the  violation of  higher
sum-rules  allows  for  superluminal  signaling,  essentially  because
probability measures can be contextual in such theories.
\end{abstract}

\maketitle

\section{Introduction}

It  is  an   interesting  and  basic  question   what  determines  the
mathematical  structure of  quantum  mechanics  (QM).  Answering  this
potentially holds a  key to resolving one of the  top open problems in
modern theoretical physics: namely, the unification of QM with general
relativity for  creating a unified  theory of quantum  gravity.  There
have been three  broad approaches to answering this  question.  In one
approach,  the framework  of  generalized  probability or  correlation
theories \cite{gpt,mag06} broader  than QM is employed,  and one tries
to identify  axioms which  would allow  us to  derive QM.   In another
approach,   one    attempts   to   identify    information   theoretic
(cryptographic)  axioms for  QM  \cite{Fu01,Smo05,CBH03}.  Finally,  a
third  approach  is to  consider  slight  variations  to QM,  such  as
modifying  the  Born  probability  rule  \cite{wei89}  or  introducing
nonunitary  evolution for  closed systems  \cite{scoar04}, and  noting
that such variations seem to  lead to implausible consequences, mainly
the  violation  of  the no-signaling  principle  \cite{gis90,  polchi,
  Her82, Gis98,  GKR99, Pat00, PB03,  Kho0, G0, GR0} or  the efficient
solution of computationally hard problems \cite{scoar04,srik10}.
   
In  an interesting  instance  of  the third  approach,  due to  Sorkin
\cite{sorkin},  quantum  measurement  statistics is  placed  within  a
hierarchy  of  probability  measure  theories, whose  members  can  be
distinguished operationally  using a generalization of  Young's double
slit experiment \cite{sorkin}.  The Sorkin  architecture is one of the
popular theories  which looks  at alternative formulations  of quantum
mechanics.   It is  actively under  study  and there  are many  recent
investigations based  on this  architecture which  involve theoretical
explorations \cite{UBE11, Nie13, caslav} as well as experimental tests
\cite{urbasi,PML12,weihs12}.  Whereas  certain generalized probability
or   correlation  theories,   such   as  those   discussed  in   Refs.
\cite{gpt,mag06},  are explicitly  designed to  be non-signaling,  the
status of signaling in Sorkin's formalism has not been studied, as far
as we know. One can  construct non-signaling toy theories that violate
the Born rule, but the  response of Refs.  \cite{gis90,polchi} to Ref.
\cite{wei89},  and  the  results  of   the  type  discussed  in  Refs.
\cite{scoar04,srik10},  suggest that  theories otherwise  close to  or
identical with QM  lead to signaling.  This therefore  motivates us to
check the  status of  signaling in Sorkin's  hierarchy, and  forms the
focus of our study in this work.

No-signaling is a  fundamental feature of QM as  we understand it, and
implies that a signal does not travel from one point to another except
through the  physical communication  of a particle.   It is in  fact a
feature  of  non-relativistic  QM,  but  compatible  with  relativity.
Einstein-Podolsky-Rosen       (EPR)      correlations      interpreted
``realistically''  imply  a nonlocal  influence,  demonstrated by  the
violation of Bell-type  inequalities \cite{chsh}, but this nonlocality
cannot be used for signaling.  Formally, no-signaling is the statement
that the reduced  density operator of a system  is unaffected by local
operations on another system with  which it may be entangled.  General
correlations  (not necessarily quantum)  over $N$  spatially separated
parties,    given     by    $P_N(o_1,\cdots,o_N|m_1,\cdots,m_N)$    is
non-signaling if:
\begin{equation}
P_{N-T}\left(o_{T+1},    \cdots,o_{N}|m_{T+1},   \cdots,m_N\right)   =
\sum_{o_1,\cdots,o_T}                  P_N                  \left(o_1,
\cdots,o_N|m_{1},\cdots,m_N\right),
\label{eq:nosig0}
\end{equation}
where $o_j$ are the outcomes and $m_j$'s are the measurements, for any
bi-partition of the $N$ into $T$ and $N-T$ parties.  

It  is  proven  in  Ref.   \cite{AAC+10} that  for  any  non-signaling
correlations $P_N$,  a necessary and  sufficient condition is  that it
can be written in a quantumlike form:
\begin{equation}
P_N(o_1,\cdots,o_N|m_1,\cdots,                  m_N)                 =
\textrm{Tr}\left(\left[\Pi^{m_1}_{o_1}    \otimes    \cdots    \otimes
  \Pi^{m_N}_{o_N}\right] \sigma\right),
\label{eq:AAC+10}
\end{equation}
where    $\sigma$   is    an   operator    such    that   $\textrm{Tr}
\left(\sigma\right) = 1$, and $\Pi^{m_j}_{o_j}$ are positive operators
in  the local  space $j$,  satisfying for  all $m_j$  the completeness
condition $\sum_{o_j}  \Pi^{m_j}_{o_j} = \mathbb{I}_j$.   If, further,
$\sigma$ is positive, the correlations  $P_N$ are quantum. If and only
if $\sigma$  corresponds to a product  state is $P_N$  local.  In this
way, this unified  framework allows us to obtain  different classes of
theories by varying the properties of $\sigma$.  If all possible local
measurements   $\Pi^{m_j}_{o_j}$,   the  method   to   extend  it   to
multi-partite systems, and the  operators $\sigma$ allowed in a theory
are  known  and  well  characterized,  then the  question  of  whether
no-signaling  is  valid in  the  theory  can  be readily  tested.   As
clarified  below, this  does not  seem to  be the  case in  the Sorkin
architecture, and we must use another approach.

The  remaining   article  is   structured  as  follows.    In  Section
\ref{sec:new}, we provide the main motivation to expect a violation of
no-signaling in the Sorkin architecture, basically hinging on the fact
that  it  is  incompatible  with non-contextuality  of  probabilities.
After presenting  in Section \ref{sec:sorq} Sorkin's  idea briefly, in
Section  \ref{sec:B2A} we propose  on its  basis a  signaling protocol
using a  non-maximally entangled state,  in which the observer  at the
interferometer   signals  another   observer   sharing  an   entangled
particle. We point  out a subtlety that can thwart  the signal in this
case, which paves  the way for the signaling  protocol proposed in the
next Section.  We propose in section \ref{sec:A2B} a rigorous argument
for the incompatibility of  no-signaling with the Sorkin architecture,
based on a  set-up wherein the interferometer is  the signal receiver,
rather than sender.  We then conclude in the next section, summarizing
the deeper implication of our work.

\section{Signaling via contextuality of probabilities \label{sec:new}}

In the foundations of quantum mechanics, we may distinguish between on
the one hand the \textit{contextuality} of hidden variable assignments
to   explain  the   outcomes  of   incompatible   measurements,  which
generalizes  quantum  nonlocality, and  is  related  to  the Bell  and
Kochen-Specker  theorems  \cite{AB11},  and  on the  other  hand,  the
\textit{non-contextuality}  of   probability  assignments  to  explain
outcomes   of   compatible   measurements,   which   generalizes   the
no-signaling principle to the no-disturbance principle, and is related
to Gleason's celebrated theorem \cite{Gle57}.

By this  theorem, assuming  the state space  to be that  of (standard)
quantum mechanics, i.e., Hilbert space, non-Bornian probabilities must
in general  be contextual.  That  this contextuality can be  the basis
for nonlocal signaling  using a 3-dimensional \textit{single-particle}
system,  was  shown by  Peres  \cite{Per93}.   Suppose  a particle  is
described by Hilbert space ${\cal H}_3 \equiv \textrm{span}(|0\rangle,
|1\rangle, |2\rangle,  |3\rangle)$, where $|j\rangle$  are 4 spatially
separated wave packets.   Further, suppose that Alice and  Bob are two
observers  who are  separated spatially.   Qutrits are  prepared  in a
source in state $|\phi\rangle  = \sum_k \alpha_k |k\rangle$ ($k=0,1,2,
3$ such  that $\sum_k |\alpha_k|^2=1$).   They are shot towards  a lab
located centrally between Alice and Bob where a beam-splitter deflects
the $|0\rangle$ spin towards Bob, while the $|1\rangle, |2\rangle$ and
$|3\rangle$ are deflected towards  Alice.  Alice's Hilbert subspace is
thus    $\textrm{span}(|1\rangle,    |2\rangle,    |3\rangle)$.     If
probabilities were  contextual (in the  sense of Gleason),  this would
mean  that   Bob's  probability  to  observe   $|0\rangle$  (which  is
$|\alpha_0|^2$ in  the standard theory) may be  different depending on
whether Alice measures in the basis $\mathcal{B}_1 \equiv \{|1\rangle,
|2\rangle,    |3\rangle\}$     or    (say)    $\mathcal{B}_2    \equiv
\{\frac{1}{\sqrt{2}}  (|1\rangle \pm  |2\rangle),  |3\rangle\}$ basis,
thereby leading to a signal.

To study the  status of signaling in the  Sorkin architecture, we will
use methods based on the above idea. However, it turns out that Peres'
protocol cannot be directly used to expose signaling, but must instead
be extended to the multi-particle case.  The reason is the possibility
of what we call \textit{local redistribution of probabilities}. In the
above example, Alice may measure  in the basis $\mathcal{B}_1$ or pass
her particles through a 3-slit  before measurement in a `screen basis'
(which can  be modelled  by a discrete  quantum Fourier  transform, as
shown   below)  $\mathcal{B}_3   \equiv   \{|s_1\rangle,  |s_2\rangle,
|s_3\rangle\}$.  Sorkin's recipe (as detailed below) only entails that
the probability $P^{\rm  Sor}(s_j)$ for Alice to find  the particle at
position  $s_j$ on  the  screen  will not  coincide  with the  quantum
mechanical   probability  $\textrm{Tr}(\rho|s_j\rangle\langle  s_j|)$.
Yet,  it may  still be  true that  $\sum_j P^{\rm  Sor}(s_j)  = \sum_j
\textrm{Tr}  (\rho|s_j\rangle\langle  s_j|)$,  so  that no  signal  is
detected by Bob in this particular experimental setting.  Effectively,
this is because  the departure from Born rule is  such that the screen
probabilities  are simply re-distributed  amongst themselves,  with no
overall modification of the probability mass in that subspace.  In the
Appendix,   we  provide   a  general   argument  for   how  contextual
probabilities can lead  to signaling across the two  parties who share
an entangled  state.  In Sections \ref{sec:B2A}  and \ref{sec:A2B}, we
provide specific thought experiments that demonstrate this signaling.

It  is important  to  stress  that our  above  argument for  signaling
presumes  the  Hilbert  space  structure  of the  state  space.   Ref.
\cite{Son14} presents a construction of non-signaling probabilities in
a  Banach space, in  particular $L^p$  spaces with  the $p$-norm  $p =
1,2,\cdots$,  which corresponds  to Hilbert  space for  $p=2$  and the
Popescu-Rohrlich  box  statistics  \cite{PR94} for  $p=\infty$.   When
$p\ne2$,  Gleason's  theorem  does  not apply,  and  thus  non-Bornian
recipes for probability do  not necessarily entail contextuality.  The
key observation here regarding the Sorkin architecture is that quantum
mechanical predictions are expected to hold to arbitrary accuracy when
the  1-slit  and  2-slit   interferometric  set-ups  are  used.   This
constrains  the state  space to  the usual  2-norm,  thereby rendering
Gleason's  theorem binding.  Our  result basically  says that  if Born
rule holds in the 2-slit context,  then there is no freedom to deviate
from this rule when multi-slit contexts are considered.
 
The   remaining  article   is   arranged  as   follows.   In   Section
\ref{sec:sorq},   we  briefly  introduce   the  Sorkin's   theory  for
generalizing quantum measure.  Based on Peres' above idea of using the
contextuality of probabilities to violate no-signaling, we present two
specific realizations of nonlocal signaling in the Sorkin architecture
in Section  \ref{sec:B2A} using a  non-maximally entangled state  in a
sufficiently high-dimensional Hilbert space.  Our result does not rule
out  modifications to  QM along  the  lines envisaged  by Sorkin,  but
suggests  regimes where  such  effects may  be  relevant.  We  briefly
adumbrate this point and conclude in Section \ref{sec:conclu}.

\section{Sorkin's generalization of quantum measure theory
\label{sec:sorq}}

In  such  an  experiment  as   envisaged  by  Sorkin,  one  assigns  a
probability measure to a set of pathways belonging to a particle being
detected at  a given point on  the screen.  For  example, consider the
double slit experiment,  with slits $A$ and $B$, in  which one or both
slits may  be left open.   For any point  on the screen, we  can write
down the  three quantities $P(A  \wedge B), P(A),  P(B)$, representing
the probability  of detection  with both slits  being open,  with only
slit  $A$  open and  with  only  slit  $B$ open,  respectively.   Here
`$\wedge$'  is   the  Boolean  \texttt{AND}   operator.   For  quantum
probability, the interference term:
\begin{equation}
I_2(A,B) \equiv P(A \wedge B) - P(A) - P(B)
\label{eq:2slits}
\end{equation}
is   non-vanishing,   i.e.,  the   2-sum   rule  $I_2(A,B)=0$,   fails
\cite{sorkin}, meaning that  probabilities with individual slits being
open are not additive.

On  the other  hand, the  quantum mechanical  Born rule  satisfies the
3-sum rule, in that the three-term interference
\begin{eqnarray}
I_3(A,B,C) \equiv P(A \wedge B \wedge  C) - P(A \wedge B) - P(A \wedge
C) - P(B \wedge C) + P(A) + P(B) + P(C)
 \label{eq:3int}
\end{eqnarray}
vanishes. Here $P(A \wedge B \wedge C)$ is the probability to detect a
particle at a given position, with  slits $A, B$ and $C$ open.  In QM,
suppose $\psi_j$  ($j =  A, B,  C$) is the  amplitude that  a particle
propagates from slit $j$ to point $x$ on the screen, then we have $P(A
\wedge B \wedge  C) \equiv |\psi_A + \psi_B  + \psi_C|^2$, $P(A \wedge
B)  =  |\psi_A  +  \psi_B|^2$,  etc., and  $P_A  =  |\psi_A|^2$,  etc.
Substituting   these   into  Eq.    (\ref{eq:3int}),   we  find   that
$I_3(ABC)=0$,    implying   that    third-order    (and   higher-order
interference)  are absent  in the  hierarchy of  sum-rules  defined as
follows.   Informally,  this is  because  interference occurs  through
mixing of pairs of paths and not triplets or quadruplets of paths.

The  validation of  the $N$-sum  rule  requires the  vanishing of  the
$N$-th order interference term
\begin{eqnarray}
I_N(A_1,A_2,\cdots,A_N) &\equiv&  P\left(\bigwedge_j A_j\right) - \sum
P([N-1]\textrm{-sets})  \nonumber \\  &+&   \sum  
P([N-2]\textrm{-sets}) 
 -  \cdots
(-1)^{N-1}\sum_{j=1}^N P(A_j),
\end{eqnarray}
where $\sum  P([N-1]\textrm{-sets})$ is the sum  of probabilities over
all choices of $(N-1)$  open slits, etc \cite{sorkin}.  Experiments to
date   place    a   stringent   upper    bound   on   such    a   term
\cite{urbasi,PML12,weihs12}.

It  is  of interest  to  know whether  such  modifications  to QM  can
accomodate other properties of QM, considered to be fundamental, since
if this  were not so,  then this incompatability  could be used  as an
axiomatic basis  \cite{Svt98} to  rule out higher  order interference.
Here  we prove  that,  under certain  assumptions,  such higher  order
super-quantum interferences indeed lead to superluminal signaling.  To
the best of our knowledge, the  issue of signaling has not been raised
in this  connection, which  is somewhat surprising,  considering works
cited  above \cite{gis90,  polchi, Her82,  Gis98, GKR99,  Pat00, PB03,
  scoar04,srik10}. The reason may  be that the formulation of Sorkin's
modification  in terms  of  sum rules  rather  than a  straightforward
change in the  dynamics makes the search for  states that would expose
the signaling less obvious.

In  developing   this  hierarchical  framework,   Sorkin  had  ignored
contributions from non-classical  i.e., looped paths \cite{sorkin}. In
a recent work, one of us has investigated the effect of including such
paths  in the  calculation  \cite{sawant}. We  find  that taking  into
account  non-classical paths  in a  triple slit  problem,  does indeed
generate a non-zero third order interference term even in the standard
theory,   due  to  contributions   from  certain   boundary  condition
considerations.  In a situation  where two partners share a correlated
state, in  standard quantum mechanics  this non-zero term  will remain
invariant  when   the  remote  partner   changes  her/his  measurement
settings. However, the part  of $I_3$ that receives contributions from
the new  physics \`a la Sorkin  may not be invariant  under actions of
the remote partner.   For simplicity of the narrative,  we assume that
such boundary-condition  based $I_3$ contribution  is re-calibrated to
0,  so  that the  condition  $I_3\ne0$  can  indeed be  considered  as
equivalent to signaling.

Another  assumption  is that  the  state  space  structure of  quantum
mechanics  holds  good,  with  only  the  dynamical  part  altered  to
accommodate  a new  interference recipe.   In this  context, it  is of
interest  to  note a  recent  work  \cite{caslav},  where the  authors
present a  formalism to  realize the higher-order  interferences, with
states being  represented by tensors of  correspondingly larger number
of indices.  In contrast to  our approach, their method allows for the
possibility  of  non-quantum  states.    It  will  be  interesting  to
investigate the consequence from our  signaling point of view for such
generalized theories.

By Gleason's  theorem, the probabilities  in the Sorkin  method, being
non-Bornian,  must be  contextual. However  they do  not automatically
lead to  signaling of the type  considered by Peres  for the following
reason. Let  Alice's measurement  apparatus be a  three-slit diaphragm
followed by one of the two following set-ups: a screen, or a system of
three telescopes each focused on  a different slit. If she measures in
the  `slit basis'  (by  using  the telescopic  instead  of the  screen
system), there  is no departure  from the standard  quantum mechanical
prediction, by design. However, if she measures in the `screen basis',
a  violation  of  the  3-sum  rule  (\ref{eq:3int})  can  cause  Bob's
probability  to  deviate  from  $|\alpha_0|^2$,  thereby  providing  a
Peres-like  mechanism  for  signaling.   However, a  signal  does  not
necessarily follow, since the failure  of the 3-sum rule may result in
re-distribution of probabilities in the screen plane without affecting
Bob's probabilities. We  may hope to avoid this  scenario if Alice and
Bob  were observing  two separated  but  \textit{entangled} particles,
because  the redistribution  may be  reflected on  Bob's side  via the
nonlocal correlations, thereby producing a signal. 

This   immediately  provides   a  motivation   for   extending  Peres'
single-particle idea  to a multi-partite situation,  and repeating the
above thought experiment there.  We  give examples of signaling in the
Sorkin  architecture in  the  following Section.  A general,  abstract
framework for expecting such a  signal in a bipartite entangled system
is given in Appendix \ref{sec:peres++}.

\section{Signaling in the Sorkin architecture: 
Interferometric sender 
\label{sec:B2A}}

In  Sorkin's approach,  the modified  theory must  be indistingishable
from  quantum mechanics  for all  single-slit and  2-slit experiments.
Intuitively, this  means (as  made rigorous in  Theorem \ref{thm:this}
below)  that the state  space of  the particle  is identical  with the
quantum mechanical  one, i.e., described  by the usual  2-norm Hilbert
space geometry.  This  is our first observation.  Now,  if further, we
assume that  probabilities are non-contextual,  then Gleason's theorem
compels them  to comply with the  Born rule.  Thus,  if $I_N\ne0$ (for
$N\ge3$),  probabilities  must  be  contextual, which  is  our  second
observation.  The crux of this  work is to show how this contextuality
can be used as a basis for nonlocal signaling.

By the first observation above, Alice  and Bob can be assumed to share
an  arbitrary  \textit{quantum}  state.    Suppose  that  this  is  an
entangled  state.    Bob  may  employ   a  multiple-slit  interference
experiment, and correspondingly, his  particle is of sufficiently high
dimension.  Alice's  is a spin-$1/2$ particle, which  ensures that its
measurement  statistics are unaffected  even if  there is  a deviation
from Born rule due to  the presence of higher order interference terms
in the Sorkin framework.

\subsection{Non-maximal entanglement} 
  
We may  begin by considering (maximal) entanglement  between the modes
of Bob's particle, that are assumed  to be localized at each slit, and
corresponding modes of Alice's particle.   The impasse we are met with
here is that the tight correlation will render Bob's modes incoherent,
precluding a  test of higher-order  interference.  On the  other hand,
making  Bob's  modes fully  coherent  renders  them disentangled  from
Alice's  ones, and  hence Alice  powerless to  remotely  prepare Bob's
ensemble.  What  is required thus is  non-maximal entanglement between
Alice's  and  Bob's  particles,  which provides  a  trade-off  between
required coherence and remote control.

The scheme below requires a system of dimensionality $2 \times 4$.  We
present a  more general version for a  $2 \times N$ $(N  > 3)$ system.
Charlie creates a non-maximally entangled state between a qubit and an
$N{+}1$-dimensional particle, of the form:
\begin{equation}
|\Psi\rangle_{AB} = \sum_{j=0}^{N-1} \alpha_j|0\rangle_A|j\rangle_B
+ \alpha_N|1\rangle_A|N\rangle_B.
\label{eq:gensig}
\end{equation}
where   $\{|j\rangle\}$  constitutes   modes  that   are  sufficiently
localized   in  the   transverse  direction.    In  this   state,  the
entanglement is such  that the first $N$ modes  are coherent with each
other  (the  off-diagonal  terms  are  non-vanishing  in  the  density
operator  when represented  in this  basis),  while the  last mode  is
incoherent from them.

\begin{figure}
\begin{center}
\includegraphics[width=8cm]{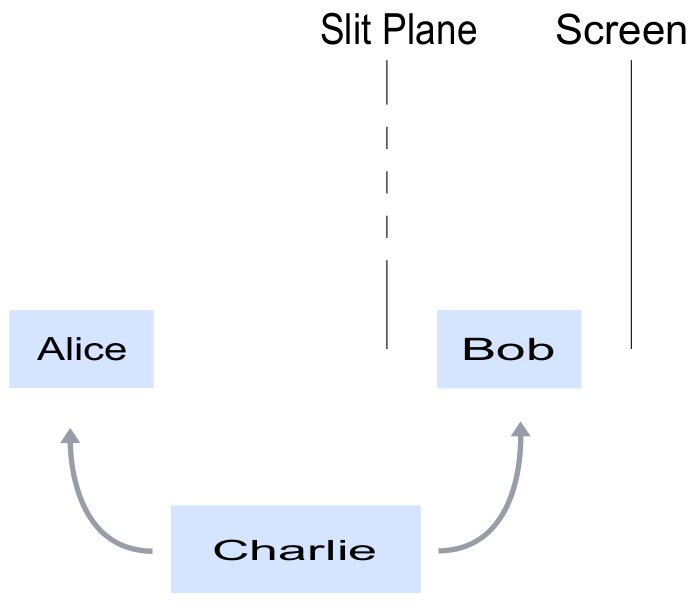}
\caption{Charlie  distributes entanglement (\ref{eq:gensig})  to Alice
  and Bob, who performs  a multi-slit interference experiment.  In the
  scheme of Section  \ref{sec:B2A}, the number of slits  $N \ge 4$ and
  Alice observes her  particle, while Bob measures his  in the slit or
  screen basis. In the scheme  of Section \ref{sec:A2B}, the number of
  slits $N \ge  3$ and Bob observes his particle  always in the screen
  basis, while  Alice measures hers  in the computational  or Hadamard
  basis.}
\label{fig:4slits}
\end{center}
\end{figure}

Charlie now distributes  the entanglement to Alice and  Bob, such that
Alice receives the first  particle in $|\Psi\rangle_{AB}$ and Bob, who
is  spatially distant  from her,  receives the  second  particle.  The
state in Alice's station, which is the reduced density operator of the
first particle, is
\begin{eqnarray}
\rho_A &=& \left(\sum_{j=0}^{N-1}|\alpha_j|^2\right)|0\rangle\langle0|
  + |\alpha_N|^2|1\rangle\langle1| \nonumber \\
\rho_B &=& \sum_{j,k=0}^{N-1}\alpha_j\alpha^\ast_k|j\rangle\langle k|
  + |\alpha_N|^2|N\rangle\langle N|.
\label{eq:rhoa}
\end{eqnarray}
Bob's particle  is passed through  a set-up consisting of  a diaphragm
with $N+1$ slits, aligned  to receive the transversely localized modes
$|j\rangle$. We may thus regard $\{|j\rangle\}$ as the `slit basis' of
Bob (Figure  \ref{fig:4slits}). Measuring his particle  in this basis,
Bob leaves Alice's particle in the state $\rho_A^{\rm slit}=\rho_A$.

For  simplicity, the diffraction  resulting from  slit passage  may be
modelled by a discrete Fourier transform \cite{nc00}:
\begin{equation}
U^f|j\rangle = \frac{1}{\sqrt{N+1}}\sum_{k=0}^N e^{2\pi i jk/(N+1)}|k\rangle,
\label{eq:discfou}
\end{equation}
where the output basis is assumed to refer to the screen.  We find:
\begin{eqnarray}
\left(I_A \otimes U^f_B\right)|\Psi\rangle_{AB}
&=& \sum_{j=0}^{N-1} \alpha_j|0\rangle_A
\frac{1}{\sqrt{N+1}}\sum_{k=0}^N e^{2\pi ijk/(N+1)}|k\rangle_B
   + \alpha_N|1\rangle_A \frac{1}{\sqrt{N+1}}
\sum_{k=0}^N e^{2\pi iNk/(N+1)}|k\rangle_B \nonumber \\
&=& \frac{1}{\sqrt{N+1}}\sum_{k=0}^N\left[ 
    \left(\sum_{j=0}^{N-1} e^{2\pi ijk/(N+1)}\alpha_j\right)|0\rangle_A 
     + e^{2\pi iNk/(N+1)}\alpha_N|1\rangle_A\right]|k\rangle_B \nonumber \\
  &\equiv& \sum_{k=0}^N\left[Y_k|0\rangle_A + Z_k|1\rangle_A\right]|k\rangle_B
 \equiv \sum_{k=0}^N |\psi_k\rangle_A|k\rangle_B. 
\label{eq:fringe}
\end{eqnarray}
Bob  measures  his particle  in  the  state  (\ref{eq:fringe}) in  the
`screen  basis'  $\{|k\rangle_B\}$.   The  probability with  which  he
detects each $|k\rangle_B$ determines  his fringe pattern and is given
by the norm of each mode in Eq. (\ref{eq:fringe}):
\begin{equation}
B_k               =               |||\psi_k\rangle||^2               =
\frac{1}{N+1}\left(\sum_{j,j^\prime}e^{2\pi   i(j^\prime   -   j)
  k/(N+1)} \alpha_{j^\prime}\alpha^\ast_j + |\alpha_N|^2 \right),
\label{eq:bj}
\end{equation}
from which it follows that  $\sum_{j=0}^N B_j = 1$, using the identity
$  \frac{1}{N+1}\sum_{k=0}^N e^{2\pi ijk/(N+1)}  = \delta_{j0}.   $ We
note   that   the   structure   of   the   fringe   pattern,   as   an
\textit{incoherent} sum of  the contribution from the last  slit and a
coherent contribution from the slits  0 through $N{-}1$, is due to the
entanglement (Figure \ref{fig:4slits}).

\subsection{Signaling protocol \label{sec:3more}}

Let the  state of Alice's  particle conditioned on Bob's  measuring in
the slit basis be denoted $\rho_A^{\rm slit}$.  Denoting the normalized
version     of    $|\psi_j\rangle$    by     $|\phi_j\rangle    \equiv
\frac{1}{\sqrt{B_j}}|\psi_j\rangle$,  we find  that the  state  of the
first particle,  conditioned on the  measurement of the second  in the
screen basis, is:
\begin{equation}
\rho_A^{\rm scr} =  \sum_j B_j |\phi_j\rangle_A\langle\phi_j| = \sum_j
|\psi_j\rangle_A\langle\psi_j| = \rho_A.
\label{eq:rhoscr}
\end{equation}
We  thus have  $\rho_A^{\rm scr}  = \rho_A^{\rm  slit}$, which  is the
statement of no-signaling in standard QM.

On the other  hand, if the $m$th-order interference ($3  \le m \le N$)
can occur, then the $B_j$'s  in Eq.  (\ref{eq:bj}), which are obtained
by the usual Born quadratic  formula, must be replaced by $B^\prime_j$
such that then there is  some point $j$ for which 
\begin{equation}
B_j \ne B_j^\prime,
\label{eq:ineq}
\end{equation}
while by demand of conservation:
\begin{equation}
\sum_j B_j^\prime  =  \sum_j  B_j  =  1.
\label{eq:eq}
\end{equation}
We note  that if  Ineq. (\ref{eq:ineq}) holds,  then there will  be at
least  another  such  $j$ for  which  it  holds,  in order  to  ensure
Eq. (\ref{eq:eq}).

Now $B_j$ corresponds to the intensity at point $j$ on the screen when
all three slits  are open, in other words, $I(A \cap  B \cap C) \equiv
|\psi_A  + \psi_B  +  \psi_C|^2$ in  the  standard quantum  formalism.
Similarly,  $I(A \cap  B) \equiv  |\psi_A +  \psi_B|^2$  etc. Plugging
these values into Eq. (\ref{eq:3int}) we find that:
\begin{eqnarray}
I_3^{\rm quantum}  &=& B_j  - |\psi_A+\psi_B|^2 -  |\psi_A+\psi_C|^2 -
|\psi_B+\psi_C|^2  + |\psi_A|^2  + |\psi_B|^2  +  |\psi_C|^2 \nonumber
\\ &=& |\psi_A+\psi_B+\psi_C|^2 -  |\psi_A+\psi_B|^2 - |\psi_A+\psi_C|^2 -
|\psi_B+\psi_C|^2  + |\psi_A|^2  + |\psi_B|^2  +  |\psi_C|^2 \nonumber
\\ &=&0.
\label{eq:GC}
\end{eqnarray}
Since the 2-slit probabilities are the same in the Sorkin architecture
as in quantum mechanics, we have
\begin{equation}
I_3^{\rm Sorkin}= I_3^{\rm quantum} -B_j + B_j^\prime.
\end{equation}
Thus, Ineq. (\ref{eq:ineq}) is equivalent to $I_3\ne0$.

Therefore, if $I_3\ne0$, we will have in place of Eq. (\ref{eq:rhoscr}),
\begin{eqnarray}
\rho_A^{\rm scr\prime} &=& 
\sum_j B^\prime_j |\phi_j\rangle_A\langle\phi_j| 
\label{siga}  = \sum_j
\frac{B_j^\prime }{B_j}|\psi_j\rangle_A\langle\psi_j| \ne \rho_A,
\label{eq:rhoscrp}
\end{eqnarray}
imply  nonlocal signaling.   By construction,  Bob's  density operator
$\rho^{\rm  scr\prime}_A$   is  normalized,  though   not  necessarily
linearly  related to  $\rho_A$. We  note that  $\rho_A^{\rm  slit}$ is
unaffected even if $I_N\ne0$ ($N  \ge 3$), since during measurement in
the  slit basis,  only a  single path  (and not  three or  more paths)
contributes to  each possible detection.  Thus, the  usual Born recipe
for calculating probabilities will hold good.

\subsection{Unitary equivalence of ensembles
\label{sec:ue}}

Let  us stress that  we only  have freedom  to modify  the probability
rule, but not  the state space.  Suppose we allow  that in addition to
$B_j$, which is replaced by  $B_j^\prime$ is the modified theory, that
the   projected   state  $|\phi_j\rangle_A$   is   also  replaced   by
$|\phi_j^\prime\rangle_A$           such           that          $\sum
B^\prime_j|\phi^\prime_j\rangle_A\langle\phi^\prime_j|.   = \sum_j B_J
|\psi_j\rangle\langle \psi_j|$, so that no-signaling is guaranteed.

By the  Hughston-Jozsa-Wootters theorem \cite{HJW},  Bob's measurement
is  equivalent to  measuring in  the rotated  basis $U_{jk}|k\rangle$,
where
\begin{equation}
\sqrt{B^\prime_j}|\phi^\prime_j\rangle            =            \sum_k
U_{jk}\sqrt{B_k}|\phi_k\rangle,
\label{eq:HJW}
\end{equation}
followed  by   standard  quantum   measurement,  as  seen   by  direct
substitution.   That  $U$ defined  in  Eq.   (\ref{eq:HJW}) is  indeed
unitary may be verified by taking the norm of the r.h.s, and verifying
that  it  yields 1  when  summed  over $j$,  only  when  $U$ has  this
property. In other  words, the new physics would  simply correspond to
Bob measuring in a rotated basis, so that we still would remain within
the scope  of the Born  rule and $I_3$  must be 0. Thus  if $I_3\ne0$,
then  the two  ensembles--  the Bornian  and  non-Bornian-- cannot  be
unitarily equivalent, and a nonlocal signal will arise.

\subsection{Local redistribution of probabilities 
\label{sec:lr}}
 
Bob can  thus potentially transmit  a superluminal signal to  Alice by
remotely preparing the state  $\rho_A$ or $\rho_A^{\rm scr\prime}$, by
measuring in the slit basis or the screen basis. However, it turns out
that the conspiracy of local redistribution of probability may nullify
this signal. We  have that $B_k = |Y_k|^2 +  |Z_k|^2$. Suppose we have
that:
\begin{equation}
B_{k}^\prime   =   |Y_k^\prime|^2   +   |Z_k|^2   \textrm{~such~that~}
\sum_{k=0}^{N-1} |Y_k|^2 = \sum_{k=0}^{N-1} |Y_k^\prime|^2.
\label{eq:bk}
\end{equation}
In  other words,  the probabilities  of $|k\rangle$'s  correlated with
$|0\rangle_A$  are  redistributed  amongst  themselves,  so  that  the
probability  for Alice  to observe  $|0\rangle_A$ or  $|1\rangle_A$ is
unaltered.

\section{Signaling with an interferometric receiver \label{sec:A2B}}

As noted  above, the nullification  of the above signal  through local
redistribution can be attributed to the interferometric observer (Bob)
choosing the  ensemble.  To  amend this, we  must have Alice  make the
choice, and Bob observe any  given point on his screen.  The situation
can be concisely presented as the following result.

\begin{thm}
Suppose  $\mathcal{T}$  is a  theory  that  is indistinguishable  from
quantum mechanics  for 1-slit and 2-slit  interferometric set-ups, but
$I_3\ne0$, then $\mathcal{T}$ is nonlocally signaling.
\label{thm:this}
\end{thm}
\textbf{Proof.}   For  simplicity, we  consider  a discrete  $N$-level
system.      Consider    a     quantum    state     $|\Psi\rangle    =
\sum_{j=1}^N\alpha_j|x_j\rangle$.  There is a state $\Psi_\mathcal{T}$
in  $\mathcal{T}$ that is  indistinghishable from  $|\Psi\rangle$ when
measurements   are  performed   in  the   basis   $\mathcal{B}  \equiv
\{|x_j\rangle\}$.   A double-slit  experiment can  be considered  as a
measurement following a rotation  in a 2-dimensional subspace. Suppose
$\mathcal{B}^\prime$  is  a   basis  obtained  from  $\mathcal{B}$  by
applying  a  two-level operation  $U_{jk}  \in  \textrm{U}(2)$ on  the
subspace  spanned by  basis elements  $x_j$ and  $x_k$  (with identity
applied on the remaining  $N-2$ dimensions).  Physically $U_{jk}$, and
thus   $\mathcal{B}^\prime$,   can   be   realized  using   a   biased
beam-splitter with a phase gate  at an input port. A 2-slit experiment
would  be  realized  using   a  Mach-Zehnder  set-up  built  from  two
beam-splitters.  By  the assumption in the  Theorem, quantum mechanics
and  theory  $\mathcal{T}$   would  be  indistinguishable  even  under
measurements  in the basis  $\mathcal{B}^\prime$. Recursively,  we can
construct other  bases using only  U(2) operations, and  the resulting
measurements cannot distinguish quantum mechanics from $\mathcal{T}$.

By  the  Reck-Zeilinger-Bernstein-Bertani  theorem \cite{RZB+94},  any
$N$-dimensional  unitary  $\mathcal{U}_N  \in  \textrm{U}(N)$  can  be
decomposed    into     at    most    $\left(     \begin{array}{c}    n
  \\ 2\end{array}\right)$ U(2) operations:
\begin{equation}
\mathcal{U}_N = U_{N,N-1}U_{N,N-2}\cdots U_{2,1} D,
\label{eq:RZB+94}
\end{equation}
where $U_{j,k}$  is general beam splitter operation  on dimensions $j$
and $k$, and an indentity operation on the remaining $N-2$ dimensions;
$D$  is a  diagonal  matrix with  entries  given by  elements of  unit
modulus.  The  implication of Eq.  (\ref{eq:RZB+94}) here  is that any
discrete Hermitian  matrix can  be measured on  a quantum  state using
only beam-splitters,  mirrors and phase gates, and  by assumption, the
result  should  not  be  able  to distinguish  quantum  mechanics  and
$\mathcal{T}$.   This   in  turn  means   that  the  state   space  of
$\mathcal{T}$  is  just the  usual  Hilbert  space  equipped with  the
2-norm.  Gleason's theorem then entails that the Born rule is the only
possibility, if probabilities are non-contextual.

If  $\mathcal{T}$ is  different  from quantum  mechanics  at third  or
higher order, then there  is an interferometric experiment involving a
three-port  beam-splitter (tritter)  or  a higher-port  beam-splitter,
where we would find $I_3\ne0$ or $I_N\ne0$ for some other higher-order
interference.  By  Gleason's theorem, such a departure  from Born rule
would  imply contextuality  which,  using the  ideas  of the  previous
subsection, we turn into a nonlocal signaling scenario.

In the modified theory $\mathcal{T}$,  suppose Alice and Bob share the
state   that  is   indistinguishable  from   the  quantum   state  Eq.
(\ref{eq:fringe}) for 1-slit and 2-slit measurements. For convenience,
we represent it as a quantum state. Alice measures her particle either
in the  computational basis or Hadamard  basis $\{|{\pm}\rangle \equiv
\frac{1}{\sqrt{2}}(|0\rangle  \pm |1\rangle)$.   In  the latter  case,
their state can be represented:
\begin{eqnarray}
|\Psi^\prime\rangle_{AB} &=& \frac{1}{\sqrt{2}}\sum_{k=0}^N
\left[ \left(Y_k + Z_k\right)|{+}\rangle_A + 
  \left(Y_k - Z_k\right)|{-}\rangle_A\right]|k\rangle_B \nonumber \\
&=& \frac{1}{\sqrt{2(N+1)}}\sum_{k=0}^N\left[ 
    \left(\sum_{j=0}^{N} e^{\frac{2\pi ijk}{N+1}}\alpha_j\right)
    |{+}\rangle_A +  
      \left(\sum_{j=0}^{N-1} e^{\frac{2\pi ijk}{N+1}}\alpha_j-
  e^{\frac{2\pi iNk}{N+1}}\alpha_N
\right)|{-}\rangle_A\right]|k\rangle_B.
\label{eq:fringez}
\end{eqnarray}
This entails an incoherent  sum of two $(N+1)$-path interference terms
as  observed  by   Bob.   By  contrast,  if  Alice   measures  in  the
computational basis, then from Eq. (\ref{eq:fringe}), we find that the
screen probability  is an incoherent  sum of an  $N$-path interference
pattern  and a  singleton  contribution.  The  Hughston-Wootters-Jozsa
theorem (Section  \ref{sec:ue}) says that  these two ensembles  can be
equivalent  and   thus  non-signaling  only  if   they  are  unitarily
equivalent, i.e., Eq. (\ref{eq:HJW}) holds. In turn, this enforces the
Born rule and vanishing $I_3$.  Any departure from the Born rule must,
under the considered assumptions, thus lead in general to signaling (A
signaling protocol  using a $3  \times 2$ dimensional system  is given
below).  \hfill $\blacksquare$

To illustrate  Theorem \ref{thm:this} with an  example, consider Alice
and Bob sharing the state
\begin{equation}
|\chi\rangle        =       |0\rangle_A\left(\alpha|0\rangle_B       +
\beta|1\rangle_B\right) + \gamma|1\rangle_A|2\rangle_B
\end{equation}
that lives in a $3  \times 2$ dimensional composite Hilbert space.  As
before, the  action of Bob's  triple-slit is modelled by  the discrete
Fourier transform
\begin{equation}
F = \frac{1}{\sqrt{3}}\left( \begin{array}{ccc}
    1 & 1 & 1 \\
    1 & e^{i\omega} & e^{-i\omega} \\
    1 & e^{-i\omega} & e^{i\omega}
\end{array}\right),
\end{equation}
where $\omega = 2\pi/3$. His particle's reduced density operator
at the screen is given by
\begin{equation}
\sigma  =  \frac{1}{3}\left(|\alpha  +  \beta|^2  +  |\gamma|^2\right)
|0\rangle_B\langle0|    +    (|\alpha    +    e^{i\omega}\beta|^2    +
|\gamma|^2)|1\rangle_B\langle1|  + (|\alpha  +  e^{-i\omega}\beta|^2 +
|\gamma|^2)|2\rangle_B\langle2|,
\label{eq:rhoh}
\end{equation}
which is  an incoherent sum  of amplitude contribution from  the first
two slits and the third.  State $\sigma$ will remain unaltered even if
the 3-sum rule  is violated, because in the  two incoherent sectors of
$|\chi\rangle$  (that  correlated  with  $|0\rangle_A$ and  that  with
$|1\rangle_A$), at most only 2 paths are available for interference.

On  the  other  hand,  under Alice's  Hadamard  transformation,  state
$|\chi\rangle$ transforms to:
\begin{equation}
|\chi^\prime\rangle        =       
\frac{|0\rangle_A}{\sqrt{2}}
\left(\alpha|0\rangle_B       + \beta|1\rangle_B + \gamma|0\rangle\right) + 
\frac{|1\rangle_A}{\sqrt{2}}
\left(\alpha|0\rangle_B       + \beta|1\rangle_B - \gamma|0\rangle\right).
\label{eq:hada}
\end{equation}
Assuming violation of the 3-sum rule, here we have the possibility for
3-path interference  in each incoherent sector.  When  particle $B$ is
subjected to the triple-slit, the joint state is:
\begin{eqnarray}
|\chi^\prime_F\rangle         &=&        \frac{1}{\sqrt{2}}|0\rangle_A
\left(\left[\frac{\alpha   +  \beta  +   \gamma}{\sqrt{3}}\right]|0\rangle_B  +   
 \left[\frac{\alpha  +
  e^{i\omega}\beta   +  e^{-i\omega}\gamma}{\sqrt{3}}\right]|1\rangle_B  
 +   \left[\frac{\alpha  +
  e^{-i\omega}\beta    +   e^{i\omega}\gamma}{\sqrt{3}}\right]|2\rangle_A\right)   
\nonumber    \\   &+&
\frac{1}{\sqrt{2}}|1\rangle_A \left(\left[\frac{\alpha +  \beta - \gamma}{\sqrt{3}}\right]|0\rangle_B +
\left[\frac{\alpha   +  e^{i\omega}\beta   -   e^{-i\omega}\gamma}{\sqrt{3}}\right]|1\rangle_B  +
\left[\frac{\alpha + e^{-i\omega}\beta - e^{i\omega}\gamma}{\sqrt{3}}\right]|2\rangle_A\right)
\label{eq:qe}
\end{eqnarray} 
Because of violation  of the 3-sum rule, the  probability to detect at
each point  on the screen is  not necessarily given  as the incoherent
sum of  the squared law term but  as the incoherent sum  of some other
function $F$, $G$, etc.,  of the amplitude contributions received from
the three slits (in as much  as the state space remains quantum).  

For  example,  in  Eq.   (\ref{eq:qe})  the  probability  for  outcome
$|0\rangle_A|0\rangle_B$  will not  be $\frac{1}{6}|\alpha  +  \beta +
\gamma|^2$  but  some  other  function  $F$  of  this  amplitude  sum;
likewise,  with the  probability  to obtain  $|1\rangle_A|0\rangle_B$.
Thus, for  outcome $|0\rangle_B\langle0|$ on  the screen, no-signaling
requires $F$ and $G$ such that
\begin{equation}
\frac{1}{2}|\alpha+\beta+\gamma|^2 + \frac{1}{2}|\alpha+\beta-\gamma|^2 =
F\left(\alpha+\beta+\gamma\right) +
G\left(\alpha+\beta-\gamma\right),
\label{eq:prescri}
\end{equation}
where  the  l.h.s  is  just  the  $\frac{1}{3}(|\alpha  +  \beta|^2  +
|\gamma|^2)$    coefficient   of    $|0\rangle_B\langle0|$    in   Eq.
(\ref{eq:rhoh}).  For arbitrary $\alpha, \beta, \gamma$, obviously the
only prescription that achieves this equality is the form given in the
l.h.s, which is the usual quadratic Born recipe. Any other rule cannot
guarantee this  equality to  hold in general  (in spite  of conserving
probability)  and will  produce  a noticeable  deviation, which  would
constitute a signal.

\section{Discussions and conclusions \label{sec:conclu}}

We have  shown that  modification to  the Born rule  \'a la  Sorkin is
incompatible  with no-signaling.   The underlying  reason is  that the
assumed validity of the Born rule in the 1-slit and 2-slit cases makes
the  relevant  state  space  to   be  identical  to  that  of  quantum
mechanics. Thus one can apply Gleason's theorem, from which it follows
that   the   violation  of   the   3-sum   rule  makes   probabilities
contextual. This does not necessarily  lead to signaling at the single
particle level, because Sorkin does not explicitly specify the form of
such generalized measures, and a local redistribution of probabilities
can potentially thwart the signal.  However, in the bipartite case, we
can arrange  so that  local redistribution is  no longer a  barrier to
signaling  under  violation  of   higher-sum  rules.   Thus  the  main
contribution of this work is to recognize the contextuality implied by
the Sorkin  architecture and to  show how to  turn it into  a nonlocal
signal.

Nevertheless, it is worth pointing out  that our result does not rule out
violation of higher sum rules, but constrains the scale of validity of
such modifications to standard QM.  They may be relevant, for example,
at Planck scales, where a break-down in Lorentz invariance is expected
because of quantum gravity considerations.  \color{black}


\appendix

\section{Extending Peres' signaling protocol to bi-partite
systems \label{sec:peres++}}

Consider a composite system $S$  described by the Hilbert space ${\cal
  H}_S  \equiv  {\cal  H}_A  \otimes  {\cal H}_B$.   A  local  unitary
operation on side $A$ has the form $\epsilon \equiv \upsilon_A \otimes
\mathbb{I}_B$,  where  $\mathbb{I}_B$  is  the identity  operation  in
${\cal H}_B$.  And similarly $\epsilon^\prime \equiv \upsilon_A^\prime
\otimes \mathbb{I}_B$.  We form the partition:
\begin{equation}
{\cal  H}_S   =  \bigoplus_k  {\cal  H}_A\otimes   {\cal  B}_k  \equiv
\bigoplus_j {\cal J}_k,
\label{eq:x}
\end{equation}
where ${\cal  H}_B =  \bigoplus_k {\cal B}_k$.   It follows  that each
partition ${\cal  J}_k$ is an invariant subspace  under $\epsilon$ and
under  $\epsilon^\prime$, in  the  sense that  if  a composite  system
exists   in  $\mathcal{J}_k$,   then  it   is  not   shifted   out  of
$\mathcal{J}_k$ if subjected to $\epsilon$ or $\epsilon^{\prime}$.

Let $\Phi_k \equiv  \left(\Phi^{(A)} \otimes \Phi^{(B)}\right)_j$ be a
fiducial,      separable       basis      for      the      complement
$\overline{\mathcal{J}}_k$.    According    to   the   assumption   of
non-contextuality   in  the   sense  of   Gleason   \cite{Gle57},  the
probability  measure   $\mu\left[{\cal  J}_k\right]$  associated  with
${\cal J}_k$ should be independent of whether the basis of measurement
in    $\overline{\mathcal{J}}_k$   is    chosen   to    be   $\epsilon
\left(\Phi_k\right)$ or $\epsilon^\prime \left(\Phi_k\right)$. We have
therefore
\begin{equation}
\mu\left[{\cal     J}_k|\epsilon(\Phi_k)\right]    =    \mu\left[{\cal
    J}_k|\epsilon^\prime(\Phi_k)\right]      \equiv     \mu\left[{\cal
    J}_k\right].
\label{eq:nocon}
\end{equation} 
In  other words,  the probability  associated with  $\mathcal{J}_k$ is
independent of  whether the basis  of measurement is completed  in the
complementary     state     space     by     $\epsilon(\Phi_k)$     or
$\epsilon^\prime(\Phi_k)$.  Now,
\begin{subequations}
\begin{eqnarray}
\mu\left[{\cal          J}_k|\epsilon(\Phi_k)\right]          &\equiv&
\sum_{a=1}^{\textrm{dim}_A}\sum_{b            \in            \beta_k}
\textrm{Prob}(A=a,B=b)        \equiv       \textrm{Prob}_B(k|\epsilon)
\\    \mu\left[{\cal    J}_k|\epsilon^\prime(\Phi_k)\right]   &\equiv&
\sum_{a^\prime=1}^{\textrm{dim}_A}\sum_{b         \in        \beta_k}
\textrm{Prob}(A^\prime=a^\prime,              B=b)              \equiv
\textrm{Prob}_B(k|\epsilon^\prime),
\end{eqnarray}
\label{eq:mu}
\end{subequations}
where  $A$  and $A^\prime$  are  random  variables representing  basis
elements        in       $\upsilon_A\left(\Phi^{(A)}\right)$       and
$\upsilon^\prime_A\left(\Phi^{(A)}\right)$ and $\beta_k$ is the set of
dimensions   in   $\Phi^{(B)}$   that  span   $\mathcal{B}_k$.    Here
$\textrm{Prob}_B(jk\xi)$ is the probability  for Bob to obtain outcome
$k$ in the $\xi$-context $(\xi = \epsilon, \epsilon^\prime)$.  Suppose
that probabilities were  contextual, and that we are  able to find $k$
such that
\begin{equation}
\mu\left[{\cal    J}_k|\epsilon(\Phi_k)\right]    \ne   \mu\left[{\cal
  J}_k|\epsilon^\prime(\Phi_k)\right].
\end{equation}
Then it follows from Eq. (\ref{eq:mu}) that
\begin{equation}
\textrm{Prob}_B(k|\epsilon) \ne \textrm{Prob}_B(k|\epsilon^\prime),
\end{equation}
which  represents  a  violation  of no-signaling.   The  existence  of
contextuality does not necessarily entail that we can find such $k$ in
a composite system.

However, if  $A$ and $B$ are  entangled, $A$ can  remotely steer $B$'s
ensemble.  These ensembles which  are unitarily equivalent in standard
QM may become inequivalent  when non-Bornian probabilities are allowed
in an ensemble-dependent  way, resulting in a signal.   This turns out
to be the origin of  the signaling we obtain in Sections \ref{sec:B2A}
and \ref{sec:A2B}.
 
\end{document}